\documentclass[oribibl,envcountsame]{llncs}

\usepackage{url,jb,amssymb}

\renewcommand{\.}[1]{\!#1\!}

\newcommand{\sortdef}{::=}

\newlog{\fix}{fix}

\newcommand{\tpl}[1]{\langle #1 \rangle}

\newcommand{\ra}{\rightarrow}

\newcommand{\lra}{\leftrightarrow}
\newcommand{\Ra}{\Rightarrow}
\newcommand{\La}{\Leftarrow}
\newcommand{\Lra}{\Leftrightarrow}

\newcommand{\N}{I\!\!N}
\renewcommand{\phi}{\varphi}
\renewcommand{\leq}[0]{\leqslant}

\newcommand{\Sat}[1]{{\sf Mod}(#1)}
\newcommand{\sat}{\mathrel{\sf sat}}
\newcommand{\impl}{\models}

\newcommand{\A}{{\cal A}}
\newcommand{\V}{{\cal V}}
\renewcommand{\L}{{\cal L}}
\newcommand{\NT}{{\cal N}}
\renewcommand{\O}{{\cal O}}
\newcommand{\G}{{\cal G}}
\newcommand{\SO}{{\cal S}}
\newcommand{\QQ}{{\cal Q}}
\newcommand{\Q}{Q}
\newcommand{\x}{{\vec{x}}}
\renewcommand{\a}{{\vec{a}}}

\newcommand{\calII}[2]{{\mathchoice
			{\cal #1\mbox{\footnotesize $\cal #2$}}
			{\cal #1\mbox{\footnotesize $\cal #2$}}
			{\cal #1\mbox{\tiny $\cal #2$}}
			{\cal #1\mbox{\tiny $\cal #2$}}}}

\renewcommand{\TH}{\calII TH}

\newcommand{\AX}{\calII AX}

\newcommand{\T}[3]{%
	\Ite{#1}%
		{\Ite{#2}%
			{\Ite{#3}%
				{{\cal T}_{#3,#2}(#1)}
				{{\cal T}_{#2}(#1)}
			}%
			{\Ite{#3}%
				{{\cal T}_{#3}(#1)}
				{{\cal T}(#1)}
			}%
		}%
		{\Ite{#2}%
			{\Ite{#3}%
				{{\cal T}_{#3,#2}}
				{{\cal T}_{#2}}
			}%
			{\Ite{#3}%
				{{\cal T}_{#3}}
				{{\cal T}}
			}%
		}%
	}

\newcommand{\abs}[1]{\mid\!#1\!\mid}
\newcommand{\vcl}[1]{{\sf vc}({#1})}

\newcommand{\set}[1]{\{ #1 \}}                  
\newcommand{\subst}[1]{\{ #1 \}}                

\renewcommand{\:}[4]{%
        {%
        \renewcommand{\:}[4]{%
                {%
                \renewcommand{\:}[4]{error\error}%
                \renewcommand{\j}{{##2}}%
                {##4}%
                ##1...##1%
                \renewcommand{\j}{{##3}}%
                {##4}%
                }%
        }%
        \renewcommand{\i}{{#2}}%
        {#4}%
        #1\ldots#1%
        \renewcommand{\i}{{#3}}%
        {#4}%
        }%
}

\renewcommand{\,}[3]{\:{,}{#1}{#2}{#3}}

\spnewtheorem{Example}[theorem]{Example}{\bfseries}{\rmfamily}

\newcommand{\9}[1]{\Ite{#1}{[#1]}{}}
\newcommand{\EXAMPLE}[2]    {\begin{Example}\9{#1}#2\end{Example}}
\newcommand{\COROLLARY}[2]  {\begin{corollary}\9{#1}#2\end{corollary}}

\newcommand{\THEOREM}[2]    {\begin{theorem}\9{#1}#2\end{theorem}}
\newcommand{\DEFINITION}[2] {\begin{definition}\9{#1}#2\end{definition}}
\newcommand{\PROOF}[1]      {\begin{proof}#1\end{proof}}
\newcommand{\PROOFSKETCH}[1]{\begin{proof}[sketch]#1\end{proof}}

\newcommand{\LABEL}[1]{\label{#1}}

\title{Axiomatization of Finite Algebras
}

\author{Jochen Burghardt}

\institute{GMD FIRST, Kekulestra{\ss}e 7, D--12489 Berlin,
	\email{jochen@first.gmd.de} }

\begin{document}

\maketitle

\begin{abstract}
We show that the set of all formulas in $n$ variables
valid in a finite class ${\bf A}$ of
finite algebras is always a regular tree language,
and compute a finite axiom set for ${\bf A}$.
We give a rational reconstruction of Barzdins'
\emph{liquid flow algorithm} \cite{Barzdin:Barzdin:1991}.
We show a sufficient condition for the existence of a class
${\bf A}$ of \emph{prototype algebras} for a given theory $\Theta$.
Such a set allows us to prove $\Theta \impl \phi$
simply by testing whether $\phi$ holds in ${\bf A}$.
\end{abstract}

\section{Introduction}
\label{Introduction}

Abstraction is a key issue in artificial intelligence.
In the setting of mathematical logic and model theory,
it is concerned with the relation between
\emph{concrete} algebras and \emph{abstract} statements about them in a
formal language.
For purely equational theories, the well--known
construction of an
initial model (e.g.\ \cite{Dershowitz:Jouannaud:1990}) allows
one to compute a kind of prototypical algebra for a given theory.
In the other direction (i.e.\ from concrete
algebras to theories), however,
no computable procedures are yet known.
While it is trivial to \emph{check}
whether a given formula is
valid in a given finite algebra, it is not clear how to \emph{find}
a finite
description of all valid formulas.

In 1991, Barzdin and Barzdin \cite{Barzdin:Barzdin:1991} proposed their
\emph{liquid--flow algorithm}
which takes an incompletely given finite algebra
and acquires \emph{hypotheses} about what are probable
axioms.
We give a rational reconstruction of this work that is
based on well--known algorithms on regular tree grammars.
We give a correspondence between Barzdins' notions and grammar notions,
showing that the liquid--flow algorithm in fact amounts to a combination
of classical grammar algorithms (Thm.~\ref{thm barzdin}).

The correspondence leads to synergies in both directions:
Barzdins' approach could be extended somewhat,
and a classical algorithm seems to be improvable in its time complexity
using the liquid--flow technique.

Next, we focus on a completely given algebra and show how to compute
finite descriptions of the set of all variable--bounded
formulas valid in it.
This set is described by a grammar (Thm.~\ref{8}) and by an axiom set
(Thm.~\ref{15}).

We relate our work to Birkhoff's variety theorem
\cite{Meinke:Tucker:1992},
which states that a class ${\bf A}$ of algebras
can be characterized by equational axioms only up to its variety closure
$\vcl{\bf A}$.
If ${\bf A}$ is a finite class of finite algebras
such that $\vcl{\bf A}$ is is finitely axiomatizable at all,
we can compute an equational axiom set for it (Cor.~\ref{17}).

As an application in the field of automated theorem proving,
we give a sufficient criterion for establishing whether
a class ${\bf A}$ of algebras
is a \emph{prototype} class for a given theory $\Theta$
(Cor.~\ref{20}).
If the criterion applies,
the validity of any formula $\phi$ in $n$ variables
can be decided
quickly and simply by merely testing $\phi$ in
${\bf A}$, avoiding the search space of usual theorem proving
procedures:
$\Theta \impl \phi$ if and only if
$\phi$ is satisfied in every $\A \in {\bf A}$.

Section~\ref{Definitions and Notations} recalls some
formal definitions.
In order to make this paper self--contained,
we refer well--known results on regular
tree grammars that are used in the sequel.
Section~\ref{Equational Theories of Finite Algebras}
first gives a rational reconstruction of Barzdins' liquid flow 
algorithm;
then
we show how to compute an axiom set for a finite class of finite
algebras.
In Sect.~\ref{Application to Equational Theories} 
and \ref{Application to Theorem Proving},
we
discuss the applications to Birkhoff characterizations and prototype
algebras in theorem proving, respectively.
A full version including all proofs
can be found in
\cite{Burghardt:2002a}.

\section{Definitions and Notations}
\label{Definitions and Notations}

\DEFINITION{Sorted term, substitution}{
We assume familiarity with the classic definitions of terms and
substitutions in a many--sorted framework.
Let $\SO$ be a finite set of sorts.
A signature $\Sigma$ is a set of function symbols $f$,
each of which has a fixed domain and range.
Let $\V$ be an infinite set of variables, each of a fixed sort.
For $S \in \SO$ and $V \subseteq \V$,
$\T{\Sigma}{S}{V}$ denotes
the set of all well--sorted terms of sort $S$
over $\Sigma$ and $V$;
let $\T{\Sigma}{}{V} := \bigcup_{S \in \SO} \; \T{\Sigma}{S}{V}$.
Let $sort(t)$ denote the unique sort of a term $t$.
$\sigma = \subst{\,1n{x_\i \mapsto t_\i}}$ denotes a well--sorted
substitution that maps each variable $x_i$ to the term $t_i$.
\qed
}

\DEFINITION{Algebra}{ \LABEL{algebra}
We consider w.l.o.g.\ term algebras factorized by a set of
operation--defining equations.
In this setting, a finite many--sorted algebra $\A$ of signature 
$\Sigma$
is given by a nonempty finite set $\A_S$ of constants
for each sort $S \in \SO$ and a set $E_\A$ consisting of exactly one
equation $f(\,1n{a_\i}) = a$ 
for each $f \in \Sigma$
with $f: \:{\times}1n{S_\i} \ra S$
and each $\,1n{a_\i \in \A_{S_\i}}$,
where $a \in \A_S$.
The $\A_S$ are just the \emph{domains} of $\A$ for each sort $S$,
while $E_\A$ defines the operations from $\Sigma$ on these domains.
Define
$\Sigma_\A := \Sigma \cup \bigcup_{S \in \SO} \A_S$.
We write $(=_\A)$ for the congruence relation induced by $E_\A$;
each ground term $t \in \T{\Sigma_\A}{S}{\set{}}$
equals exactly one $a \in \A_S$.
\qed
}

We will only allow closed quantified equations as formulas.
This is sufficient since an arbitrary formula can always be transformed
into prenex normal form, and we can model predicates and junctors by
functions into a dedicated sort $Bool$.

\DEFINITION{Formula, theory}{ \LABEL{def formula}
For a $k$--tuple
$\x = \tpl{ \,1k{x_\i} } \in \V^k$
such that $x_i \neq x_j$ for $i \neq j$,
define
$\QQ(\x)
:= \set{ \:{}1k{q_\i x_\i} \mid \,1k{q_\i} \in \set{ \forall, \exists}}$
as the set of all quantifier prefixes over $\vec{x}$.
Any expression of the form $\Q: t_1 =_S t_2$
for $\Q \in \QQ(\x)$ and $t_1,t_2 \in \T{\Sigma}{S}{\x}$
is called a formula over $\Sigma$ and $\x$.
We will sometimes omit the index of $(=_S)$.
We denote a formula by $\phi$, and a set of formulas, 
also called theory, by $\Theta$.
\qed
}

When encoding predicates and junctors using a sort $Bool$,
in order to obtain an appropriate semantics%
	\footnote{%
	\newcommand{\tf}{a}%
	If in some algebra $\A$
	we had $\A_{Bool} = \set{ \tf }$ 
	and $(true = \tf), (false = \tf) \in E_\A$,
	any formula $\phi$ was valid in $\A$.
	}
it is necessary and sufficient to
fix the interpretation of the sort $Bool$ accordingly
for every algebra under consideration.
Therefor, we define below the notion of an \emph{admitted algebra},
and let the definition of $\sat$, $\impl$, etc.\ depend on it.

We tacitly assume that,
\begin{itemize}
\item when we consider only equations,
	each algebra is admitted,
	while,
\item when we consider arbitrary predicates, junctors and a sort $Bool$,
	only algebras with an appropriate interpretation of $Bool$
	are admitted.
\end{itemize}

\DEFINITION{Admitted algebras}{ \LABEL{19}
Let a signature $\Sigma$ be given.
Let $\SO_{\fix} \subseteq \SO$ be a set of sorts;
and let $\Sigma_{\fix}$ be the set of all $f \in \Sigma$ that have all
argument and result sorts in $\SO_{\fix}$.
Let a fixed $\Sigma_{\fix}$--algebra $\A_{\fix}$ be given;
we denote its domain sets by $\A_{\fix,S}$.

We say that a $\Sigma$--algebra $\A$ is admitted
if $\A_S = \A_{\fix,S}$ for each $S \in \SO_{\fix}$ and
$t_1 =_\A t_2 \Lra t_1 =_{\A_{\fix}} t_2$
for all $t_1,t_2 \in \T{\Sigma_\A}{S}{\set{}}$
and $S \in \SO_{\fix}$.
\qed
}

\DEFINITION{Validity}{
For an admitted
$\Sigma$--algebra $\A$
and
a formula $\phi$,
we write
$\A \sat \phi$ if $\phi$ is valid in $\A$,
where equality symbols $(=_S)$ in $\phi$
are interpreted as identity relations on $\A_S$,
rather than by an arbitrary congruence on it.
For a class of admitted $\Sigma$--algebras ${\bf A}$,
and a theory $\Theta$,
we similarly write
${\bf A} \sat \phi$,
$\A \sat \Theta$, and
${\bf A} \sat \Theta$.
Define $\Theta_1 \impl \Theta_2$
if $\A \sat \Theta_1$ implies $\A \sat \Theta_2$ for each admitted
$\Sigma$--algebra $\A$.

If we choose $\SO_{\fix} := \set{}$, each algebra is admitted.
Choosing $\SO_{\fix} := \set{Bool}$,
$\Sigma_{\fix} := \set{(\neg), (\wedge), (\vee), (\ra), (\lra)}$,
and $\A_{\fix}$ as the two--element Boolean algebra,
we prescribe the interpretation of $Bool$ for each admitted algebra.
\qed
}

\DEFINITION{Complete theorem sets}{ \LABEL{2}
For a $\Sigma$--algebra $\A$, and a tuple $\x$ of variables as in
Def.~\ref{def formula},
define
$\TH_\x(\A) :=$
$$\set{\Q: t_1 =_S t_2
\mid \Q \in \QQ(\x), \;
S \in \SO, \;
t_1,t_2 \in \T{\Sigma}{S}{\x}, \;
\A \sat (\Q: t_1 =_S t_2)}$$
as the set of all formulas over $\Sigma$ and $\x$
that are valid in $\A$.
The elements of $\TH_\x(\A)$ can be considered as terms over the 
extended signature 
$$\Sigma \cup \set{(=_S) \mid S \in \SO} 
\cup \set{(\Q:) \mid \Q \in \QQ(\x)}.$$
For a class ${\bf A}$ of $\Sigma$--algebras, define
$\TH_\x({\bf A}) := \bigcap_{\A \in {\bf A}} \TH_\x(\A)$.
\qed
}

\EXAMPLE{}{ \LABEL{exm algebra}
The algebra $\A_2$, defined by
$\A_{Nat} := \set{0,1}$
and $E_\A = \set{0\.+0\.=0, \; 0\.+1\.=1, \; 1\.+0\.=1, \; 1\.+1\.=0}$,
is a $\Sigma$--algebra
for $\Sigma = \set{ 0, (+)}$.
The set $\TH_{\tpl{x,y}}(\A_2)$
contains the formula $\forall x \exists y: x\.+y\.=0$,
but not $\forall x \exists y: x\.+y\.=1$,
since $1 \not\in \Sigma$.
\qed
}

\DEFINITION{Regular tree grammar}{ \LABEL{height}
A regular tree grammar $\G$
consists of rules 
$N \sortdef \:{\mid}1m{ f_\i( \,{1}{n_\i}{N_{\i\j}} ) }$
or
$N \sortdef \:{\mid}1m{ N_\i }$
where $N,N_i,N_{ij}$ are nonterminal symbols
and $f_i \in \Sigma$.
Note that $n_i$ may be also $0$.
Each $f_i( \,{1}{n_i}{N_{i\i}} )$  or $N_i$
is called an alternative.
$N,N_i,N_{ij}$ are assigned a sort each that have to fit with each other
and with $f_i$.

The size $\abs{\G}$ of $\G$ is its total number of alternatives.
We denote the set of nonterminals of $\G$ by $\NT$.
The language produced by a nonterminal $N$ of $\G$ 
is denoted by $\L_\G(N)$, it is a set of ground terms over $\Sigma$;
if $N$ is the start symbol of $\G$, we also write $\L(\G)$.
$\G$ is called deterministic if no different rules have identical
alternatives.

Define the generalized height $hg(t)$ of a ground term $t$ by
$$hg(f(\,1n{t_\i})) := \max \set{ \,1n{hg(t_\i)} } + hg(f),$$
where $\max \set{} := 0$,
and $hg(f) \in \N$ may be defined arbitrarily.
For a nonterminal $N$ of a grammar $\G$,
define $hg(N)$
as the minimal height of any term in $\L_\G(N)$,
it is $\infty$ if $\L_\G(N)$ is empty.
\qed
}

\newcommand{\intern}{incorporate}
\newcommand{\lift}{lift}
\newcommand{\intersect}{intersect}
\newcommand{\unite}{unite}
\newcommand{\tagunite}{tag}

\THEOREM{Properties of regular tree grammars}{ \LABEL{properties}
\begin{enumerate}
\item \emph{Incorporating} \cite[Sect.6]{McAllester:1992}
	\\
	Given a finite many--sorted $\Sigma$--algebra $\A$,
	a grammar $\G = \intern(\A)$ of size $\abs{E_\A}$
	can be computed in time $\O(\abs{E_\A})$ such that
	$$\forall S \in \SO, a \in \A_S \;
	\exists N_a \in \NT: \;\;
	\L_\G(N_a) 
	= \set{t \in \T{\Sigma_\A}{}{\set{}} \mid t =_\A a}.$$
\item \emph{Externing}
	\cite[Sect.3]{McAllester:1992}
	\LABEL{p externing}
	\\
	Given a deterministic grammar $\G$, a set $E$ of $\abs{\G}$
	ground equations can be computed in time $\O(\abs{\G})$
	such that 
	for all ground terms $t_1,t_2$:
	$$t_1 =_E t_2 \;\Lra\;
	\exists N \in \NT: \; t_1,t_2 \in \L_\G(N),$$
	where $(=_E)$ denotes the congruence induced by $E$.
\item \emph{Lifting}
	\cite[Thm.7 in Sect.1.4]{Comon:Dauchet:Gilleron:1999}
	\\
	Given a grammar $\G$ and a ground substitution $\sigma$,
	a grammar $\G' = \lift(\G,\sigma)$
	of size $\abs{\G} + \abs{\NT} \cdot \abs{\dom \sigma}$
	can be computed in time $\O(\abs{\G'})$
	such that 
	$$\forall N \in \NT \;
	\exists N' \in \NT': \;\;
	\L_{\G'}(N')
	= \set{ t \in \T{}{}{\dom \sigma} \mid \sigma t \in \L_\G(N)},$$
	where $\NT$ and $\NT'$ is the set of nonterminals of $\G$ and
	$\G'$, respectively.
	Note that the signature gets extended by $\dom \sigma$;
	these variables are treated as constants in $\G'$.
\item \emph{Intersection}
	\cite[Sect.1.3]{Comon:Dauchet:Gilleron:1999}
	\LABEL{p intersect}
	\\
	Given $n$ grammars $\,1n{\G_\i}$,
	a grammar $\G = \intersect(\,1n{\G_\i})$
	of size $\:{\cdot}1n{\abs{\G_\i}}$
	can be computed in time $\O(\abs{\G})$
	such that for each 
	$$\forall \,1n{N_{i_\i} \in \NT_\i}
	\exists N_{\,1n{i_\i}} \in \NT: \;\;
	\L_\G(N) = \bigcap_{j=1}^n \L_{\G_i}(N_{i_j}).$$
\item \emph{Restriction}
	[special case of~\ref{p intersect}]
	\\
	Intersection of one grammar $\G_1$
	with a term universe $\T{\Sigma}{S}{V}$,
	such that
	$$\forall N_1 \in \NT_1 \exists N \in \NT: \;\;
	\L_\G(N) = \L_{\G_1}(N_1) \cap \T{\Sigma}{S}{V}$$
	can be done in time $\O(\abs{\G_1})$ by removing all symbols not
	in $\Sigma$.
\item \emph{Union}
	\cite[Sect.1.3]{Comon:Dauchet:Gilleron:1999}
	\\
	Given $n$ grammars $\,1n{\G_\i}$,
	a grammar $\G = \unite(\,1n{\G_\i})$
	of size $n+\:{+}1n{\abs{\G_\i}}$
	can be computed in time $\O(n)$
	such that
	$$\L(\G) = \bigcup_{i=1}^n \L(\G_i)$$
	by adding one rule.
\item \emph{Composition}
	[Trivial]
	\\
	Given an $n$--ary function symbol $f$,
	a grammar $\G$,
	and nonterminals $\,1n{N_\i}$,
	a grammar $\G' = \tagunite(\G,f(\,1n{N_{\i}}))$
	of size
	$\abs{\G}+1$
	can be computed in time $\O(1)$ such that
	$$\L_{\G'}(N)
	= \set{ f(\,1n{t_\i})
	\mid \bigwedge_{i=1}^n t_i \in \L_\G(N_i) }$$
	for a certain nonterminal $N$, by
	adding one rule.
\item \emph{Weight computation}
	\cite[Sect.4]{Aiken:Murphy:1991b}
	\LABEL{p weight}
	\\
	Given a grammar $\G$, the heights $hg(N)$ can be computed 
	for all nonterminals
	$N \in \NT$ simultaneously in time $\O(\abs{\NT}^2)$.
\item \emph{Language enumeration}
	\cite[Fig.21]{Burghardt:Heinz:1996}
	\LABEL{p enum}
	\\
	Given a grammar $\G$ and the heights of all nonterminals,
	the elements of $\L_\G(N)$ can be enumerated in order of
	increasing height in time linear in the sum of their sizes
	by a simple {\sc Prolog} program.
\qed
\end{enumerate}
}

\section{Equational Theories of Finite Algebras}
\label{Equational Theories of Finite Algebras}

First, we give a rational reconstruction of the \emph{liquid flow
algorithm} of
Barzdin and Barzdin \cite{Barzdin:Barzdin:1991}.
They
use labeled graphs to compute an axiom set from an incompletely given
finite algebra.
Their approach can be reformulated in terms of regular tree languages
using the correspondence of notions shown in Fig.~\ref{Barzdin
correspondence}.
Our following theorem corresponds to 
their main result, Thm.~2.
It is in fact
a slight extension, as it allows for sorts and for substitutions
that map several variables to the same value.

On the other hand,
the \emph{liquid flow algorithm} turns out to be an improvement of the
weight computation algorithm from Thm.~\ref{properties}.\ref{p weight}.
Both are fixpoint algorithms, and identical except for minor, but
important, modifications.
The algorithm from Thm.~\ref{properties}.\ref{p weight}
has a complexity of $\O(\abs{\NT}^2)$,
while Barzdins' algorithm runs in $\O(\abs{\NT})$,
exploiting the fact that always $hg(f) \leq 1$ and therefor the 
first value $< \infty$
assigned to some $hg(N)$ must be its final one already.
Since in each cycle at least one $N$ must change its assigned $hg$
value%
	\footnote{%
	Unless the fixpoint has been reached already
	}%
,
by an appropriate incremental technique \emph{(water front)\/},
linear complexity can be achieved.
A formal complexity proof of this improved grammar fixpoint algorithm,
extended to somewhat more general weight definitions,
shall appear in \cite{Burghardt:2002a}.

\THEOREM{Reconstruction of Barzdin}{ \LABEL{thm barzdin}
Given a $\Sigma$--algebra $\A$,
domain elements
$b_1,...,b_n \in \A_{S_0}, \,1k{\,1n{a_{\j\i}} \in \A_{S_\i}}$,
and defining
$\sigma_i = \subst{\,1k{x_\i \mapsto a_{i\i}}}$
for $i=\,1n\i$
and $V := \set{\,1k{x_\i}}$,
the set of terms
$T = \set{ t \in \T{\Sigma}{S_0}{V}
\mid \bigwedge_{i=1}^n \sigma_i t =_\A b_i}$
is a regular tree language.
A grammar for it can be computed in time $\O(\abs{E_\A}^n)$.
After computing nonterminal weights in time $\O(\abs{E_\A}^{2n})$,
the language
elements can be enumerated in order of increasing height in linear
time.
}
\PROOF{
Using the notions of Thm.~\ref{properties},
let
$\G_0 := \intern(\A)$,
and $\G_i := \lift(\G_0,\sigma_i)$ for $i=\,1n\i$.
We have
$$\L_{\G_i}(N_a) 
= \set{ t \in \T{\Sigma_A}{}{V} \mid \sigma_i t =_\A a}.$$
Let $\G := \intersect(\,1n{\G_\i},\T{\Sigma}{}{V})$,
then
$$\L_\G(N_{\,1n{a_\i}})
= \set{ t \in \T{\Sigma}{}{V}
\mid \bigwedge_{i=1}^n \sigma_i t =_\A a_i}.$$
Hence $T = \L_\G(N_{\,1n{b_\i}})$.
Note that $\G$ itself does not depend on $\,1n{b_\i}$.
Compute the height of all nonterminals of $\G$ using
Thm.~\ref{properties}.\ref{p weight},
using $hg(x_i) := 0$  for $x_i \in V$
and $hg(f) := 1$ for $f \in \Sigma$.
Use Thm.~\ref{properties}.\ref{p enum}
to enumerate the terms of $\L_\G(N_{\,1n{b_\i}})$.
\qed
}

\begin{figure}[htb]
\begin{center}
\newcommand{\0}{\hspace*{0.1cm}}
\newcommand{\1}[1]{\parbox[t]{2.9cm}{\raggedleft #1}}
\begin{tabular}[t]{@{}|@{\0}r@{\0}|@{\0}p{8.0cm}@{\0}|@{}}
\hline
{\bf Barzdin \cite{Barzdin:Barzdin:1991}} & {\bf Tree Grammars}     \\
\hline
\hline
sample $P$
	& equations $E_\A$ from Def.~\ref{algebra}	\\
\hline
open term, level
	& term in $\T{\Sigma}{}{\V}$, height	\\
\hline
closed term
	& term in $\T{\Sigma_\A}{}{\set{}}$	\\
\hline
sample graph
        & grammar $\G = \intern(\A)$	\\
\hline
domain node
        & nonterminal $N_a$     \\
\hline
functional node
        & expression $f(N_{a_1},\ldots,N_{a_n})$        \\
\hline
upper node
	& $N_a$,
	\hspace*{0.3cm}
	if
	$N_a \sortdef \ldots f(\,1n{N_{a_\i}}) \ldots$	\\
\hline
lower nodes
	&
	$\,1n{N_{a_\i}}$	\\
\hline
node weight
        & language height $hg(N)$ from Def.~\ref{height}  \\
\hline
chain of dotted arcs
	& rule rhs with alternatives ordered by increasing height \\
\hline
\1{annotated sample graph}
	& grammar with heights of nonterminals
	obtained from Thm.~\ref{properties}.\ref{p weight}	\\
\hline
\emph{liquid--flow} algorithm
        & (improved) height computation algorithm from
	Thm.~\ref{properties}.\ref{p weight} \\
\hline
$\alpha$--term
	& term in
	$\T{\Sigma_\A}{}{\set{}}
	\cap \bigcup_{a \in \A_S} \L(N_a)$	\\
\hline
\1{minimal $\alpha$--term of domain node $d$}
	& term $t \in \L(N_d)$ of minimal height	\\
\hline
\1{minimal $\alpha$--term of functional node}
	& term $t \in \L(f(\,1n{N_{a_\i}}))$ of minimal height	\\
\hline
Theorem 2
	& Theorem \ref{thm barzdin}	\\
\hline
Theorem 1
	& Theorem \ref{thm barzdin} for $n=1$	\\
\hline
\end{tabular}
\caption{Correspondence of notions between \cite{Barzdin:Barzdin:1991}
        and regular tree grammars}
\LABEL{Barzdin correspondence}
\end{center}
\end{figure}

Barzdin and Barzdin allow to specify an algebra incompletely,
since their main goal is to acquire \emph{hypotheses} about what are
probable axioms.

We now investigate the special case that the substitutions
$\,1n{\sigma_\i}$ in Thm.~\ref{thm barzdin} describe \emph{all} possible
assignments of algebra domain elements to the variables $\,1k{x_\i}$.
This way, we obtain \emph{certainty} about the computed axioms --
they are guaranteed to be valid in the given algebra.

\THEOREM{Computing complete theorem sets}{ \LABEL{8}
Let $\x = \tpl{\,1k{x_\i}}$ be a $k$--tuple of variables,
$\A$ be a finite $\Sigma$--algebra and 
${\bf A}$ a finite class of finite $\Sigma$--algebras,
then
$\TH_\x(\A)$ and $\TH_\x({\bf A})$ are regular tree languages.
}
\PROOFSKETCH{
Define $\A_\x := \:{\times}1k{\A_{sort(x_\i)}}$.
For each
$\a
\in \A_\x$,
let $\sigma_\a := \subst{ \,1k{x_\i \mapsto a_{\i}} }$.
Let $\G = \intern(\A)$
and $\G_\a = \intersect(\lift(\G,\sigma_\a),\T{\Sigma}{}{\x})$
for $\a \in \A_\x$.
For $S \in \SO$ and $a \in \A_S$, let
$\G_{\a,a} = \tagunite(\G_\a,N_a =_S N_a)$,
where $(=_S)$ is a new binary infix function symbol.
Let
$\G'_\a
= \unite(\set{\G_{\a,a} \mid S \in \SO, a \in \A_S})$
for each $\a \in \A_\x$.
We have
$$\L(\G'_\a)
= \set{ t_1 =_S t_2
\mid S \in \SO, \;
t_1,t_2 \in \T{\Sigma}{S}{\x}, \;
\sigma_\a t_1 =_\A \sigma_\a t_2}.$$
Now, for each $\Q \in \QQ(\x)$,
apply set operations corresponding to $\Q$
to the $\G_\a$;
e.g.\ if $k=2$ and $Q = (\forall x_1 \exists x_2)$,
let
$$\G_\Q
= \intersect(\set{
\unite(\set{\G_{a_1,a_2} \mid a_2 \in \A_{sort(a_2)}})
	\mid a_1 \in \A_{sort(a_1)}}).$$
In general, we get
$$\L(\G_\Q)
= \set{t_1 =_S t_2
\mid S \in \SO, \;
t_1,t_2 \in \T{\Sigma}{S}{\x}, \;
\A \sat (\Q: t_1 = t_2)}.$$
For each $\Q \in \QQ(\x)$,
let $\G'_\Q = \tagunite(\G_\Q,\Q: \G_\Q)$,
where $(\Q\!:)$ is a new unary prefix function symbol.
Let $\G' = \unite(\set{\G'_\Q \mid \Q \in \QQ(\x)})$,
then
$\TH_\x(\A) = \L(\G')$.
From this,
we immediately get a grammar for $\TH_\x({\bf A})$
by Thm.~\ref{properties}.\ref{p intersect}.
\qed
}

\begin{figure}
\begin{center}
\begin{tabular}[t]{@{}l*{22}{c}@{}}
$N_{0000}$ & $\sortdef$ & $0$
        & $\mid$ & $N_{0000}+N_{0000}$ & $\mid$ & $N_{0011}+N_{0011}$
        & $\mid$ & $N_{0101}+N_{0101}$ & $\mid$ & $N_{0110}+N_{0110}$ \\
$N_{0011}$ & $\sortdef$ & $x$
        & $\mid$ & $N_{0000}+N_{0011}$ & $\mid$ & $N_{0011}+N_{0000}$
        & $\mid$ & $N_{0101}+N_{0110}$ & $\mid$ & $N_{0110}+N_{0101}$ \\
$N_{0101}$ & $\sortdef$ & $y$
        & $\mid$ & $N_{0000}+N_{0101}$ & $\mid$ & $N_{0011}+N_{0110}$
        & $\mid$ & $N_{0101}+N_{0000}$ & $\mid$ & $N_{0110}+N_{0011}$ \\
$N_{0110}$ & $\sortdef$ &
        &        & $N_{0000}+N_{0110}$ & $\mid$ & $N_{0011}+N_{0101}$
        & $\mid$ & $N_{0101}+N_{0011}$ & $\mid$ & $N_{0110}+N_{0000}$ \\
$N_{0*0*}$ & $\sortdef$ &
	&        & $N_{0000}$ & $\mid$ & $N_{0001}$
	& $\mid$ & $N_{0100}$ & $\mid$ & $N_{0101}$	\\
$N_{0*1*}$ & $\sortdef$ & \multicolumn{3}{l}{$\ldots$}	\\
$N_{\forall\forall}$ & $\sortdef$ &
        &        & $N_{0000}=N_{0000}$ & $\mid$ & $N_{0011}=N_{0011}$
        & $\mid$ & $N_{0101}=N_{0101}$ & $\mid$ & $N_{0110}=N_{0110}$ \\
$N_{\forall\exists}$ & $\sortdef$ &
	&        & $N_{0*0*}=N_{0*0*}$ & $\mid$ & $N_{0*1*}=N_{0*1*}$
	& $\mid$ & $N_{0**0}=N_{0**0}$ & $\mid$ & $N_{0**1}=N_{0**1}$ \\
      &&& $\mid$ & $N_{*00*}=N_{*00*}$ & $\mid$ & $N_{*01*}=N_{*01*}$
	& $\mid$ & $N_{*10*}=N_{*10*}$ & $\mid$ & $N_{*11*}=N_{*11*}$ \\
      &&& $\mid$ & $N_{*0*0}=N_{*0*0}$ & $\mid$ & $N_{*0*1}=N_{*0*1}$
	& $\mid$ & $N_{*1*0}=N_{*1*0}$ & $\mid$ & $N_{*1*1}=N_{*1*1}$ \\
$N_{\exists\forall}$ & $\sortdef$ &
	&        & $N_{00**}=N_{00**}$ & $\mid$ & $N_{01**}=N_{01**}$ \\
      &&& $\mid$ & $N_{**00}=N_{**00}$ & $\mid$ & $N_{**01}=N_{**01}$
	& $\mid$ & $N_{**10}=N_{**10}$ & $\mid$ & $N_{**11}=N_{**11}$ \\
$N_{\exists\exists}$ & $\sortdef$ &
	&        & $N_{0***}=N_{0***}$
	& $\mid$ & $N_{*0**}=N_{*0**}$ & $\mid$ & $N_{*1**}=N_{*1**}$ \\
      &&& $\mid$ & $N_{**0*}=N_{**0*}$ & $\mid$ & $N_{**1*}=N_{**1*}$
	& $\mid$ & $N_{***0}=N_{***0}$ & $\mid$ & $N_{***1}=N_{***1}$ \\
$N$ & $\sortdef$ &
	&        & $\forall x \forall y: N_{\forall\forall}$
	& $\mid$ & $\forall x \exists y: N_{\forall\exists}$
	& $\mid$ & $\exists x \forall y: N_{\exists\forall}$
	& $\mid$ & $\exists x \exists y: N_{\exists\exists}$	\\
\end{tabular}
\caption{Grammar $\G'$ for $\TH_\x(\A)$ in Exm.~\ref{exm TH}}
\LABEL{fig exm TH}
\end{center}
\end{figure}

\begin{figure}
\begin{center}
\begin{tabular}[t]{@{}l@{\hspace*{1.0cm}}l@{}}
\begin{tabular}[t]{@{}*{8}{c}@{}}
&& $N$ 	\\
\cline{1-5}
$\forall x \forall y:$ &&&& $N_{\forall\forall}$	\\
\cline{3-7}
$\forall x \forall y:$ 
	&& \multicolumn{2}{l}{$N_{0110}$} & $=$ 
	& \multicolumn{2}{r}{$N_{0110}$}	\\
\cline{2-4}
\cline{6-8}
$\forall x \forall y:$ 
	& $N_{0011}$ & $+$ & $N_{0101}$ & $=$ 
	& $N_{0101}$ & $+$ & $N_{0011}$ \\
\cline{2-2}
\cline{4-4}
\cline{6-6}
\cline{8-8}
$\forall x \forall y:$ & $x$ & $+$ & $y$ & $=$ & $y$ & $+$ & $x$ \\
\end{tabular}
&
\begin{tabular}[t]{@{}*{6}{c}@{}}
&& $N$	\\
\cline{1-4}
$\forall x \exists y:$ &&& $N_{\forall\exists}$	\\
\cline{3-6}
$\forall x \exists y:$ && $N_{0**0}$ && $=$ & $N_{0**0}$	\\
\cline{3-3}
\cline{6-6}
$\forall x \exists y:$ && $N_{0110}$ && $=$ & $N_{0000}$	\\
\cline{2-4}
\cline{6-6}
$\forall x \exists y:$ & $N_{0011}$ & $+$ & $N_{0101}$ & $=$ & $0$ \\
\cline{2-2}
\cline{4-4}
$\forall x \exists y:$ & $x$ & $+$ & $y$ & $=$ & $0$	\\
\end{tabular}
\\
\end{tabular}
\caption{Example derivations from $\G'$ in Exm.~\ref{exm TH}}
\LABEL{fig exm deriv}
\end{center}
\end{figure}

\EXAMPLE{}{ \LABEL{exm TH}
Let us compute $\TH_\x(\A_2)$ for the algebra $\A_2$
from Exm.~\ref{exm algebra} and $\x = \tpl{x,y}$.
We have the substitutions
$\sigma_{00}, \sigma_{01}, \sigma_{10}, \sigma_{11}$
and use the naming convention
$$\L(N_{ijkl})
= \L(\G_{00,i}) \cap \L(\G_{01,j})
\cap \L(\G_{10,k}) \cap \L(\G_{11,l}),$$
e.g.\
$\L(N_{0011})
= \set{ t \in \T{\Sigma}{}{\x}
\mid \A_2 \sat (\sigma_{00} t = \sigma_{01} t = 0 \wedge
\sigma_{10} t = \sigma_{11} t = 1)}$.
A ``*'' may serve as \emph{don't care symbol},
e.g.\
$\L(N_{i*k*}) = \L(\G_{00,i}) \cap \L(\G_{10,k})$.

From Thm.~\ref{8},
after incorporating, lifting, and restricting,
we obtain e.g.\ $\G_{00}$, with nonterminals $N_{0***}$ and $N_{1***}$.
As $\L(N_{1***})$ turns out to be empty, we simply have
$$N_{0***} \sortdef 0 \mid x \mid y \mid N_{0***}+N_{0***}.$$
We obtain $\G'_{00}$ by just adding the rule
$N_= \sortdef (N_{0***} = N_{0***})$.
To compute $\G_{\forall\forall}$,
we build all intersections
$N_{ijkl}$ without ``*'';
only four of them turn out to be nonempty,
their rules are shown in Fig.~\ref{fig exm TH}.
The grammar
$\G_{\forall\forall}$ consists of these rules and an additional one
for its start symbol $N_{\forall\forall}$.
The grammars $\G_{\forall\exists}$, $\G_{\exists\forall}$, and
$\G_{\exists\exists}$ have similar starting rules, which use only
nonterminals $N_{ijkl}$ containing a ``*''.
Since the rules for the latter are trivial,
only the one for $N_{0*0*}$ is shown.
Finally, the grammar $\G'$ for $\TH_\x(\A_2)$
consists of all these rules and
an additional one for its start symbol $N$.
Figure~\ref{fig exm deriv} show some example derivations.
\qed
}

\THEOREM{Computing complete axiom sets}{ \LABEL{15}
The sets
$\TH_\x(\A)$
and
$\TH_\x({\bf A})$
obtained from Thm.~\ref{8}
can be represented as the
deductive closure of a finite set of formulas,
called
$\AX_\x(\A)$
and
$\AX_\x({\bf A})$,
respectively.
We have:
$$\begin{array}{lrcll}
\forall t_1,t_2 \in \T{\Sigma}{}{\x}:
	& \A \sat t_1 = t_2 
	& \Lra
	& \AX_\x(\A) \impl t_1 = t_2
	& \mbox{, and}	\\
\forall t_1,t_2 \in \T{\Sigma}{}{\x}:
	& {\bf A} \sat t_1 = t_2)
	& \Lra 
	& \AX_\x({\bf A}) \impl t_1 = t_2
	& .	\\
\end{array}$$
}
\PROOFSKETCH{
First, we consider the purely universal formulas in $\TH_\x(\A)$.
Using the notions of Thm.~\ref{8},
the grammar $\G_{\forall\ldots\forall}$
is deterministic since no union operations were involved in its
construction.
Using Thm.~\ref{properties}.\ref{p externing},
we get a finite set $E_{\forall\ldots\forall}$ of equations,
each of which we compose with the appropriate universal quantifier 
prefix
$(\:{}1k{\forall x_\i}:)$.
The resulting formula set $E'_{\forall\ldots\forall}$
implies any purely universal equation valid in
$\A$.
By construction of $E_{\forall\ldots\forall}$,
it can reduce
each term $t$ in any quantified equation in $\TH_\x(\A)$
to a unique normal form.
Let $NF$ denote the set of all those normal forms;
it is finite since $\abs{\NT}$ is finite.

Next, for any quantifier prefix $\Q$ containing some ``$\exists$'',
let
$$E'_\Q
:= \set{ \Q: t_{1n} = t_{2n}
\mid t_{1n}, t_{2n} \in NF,  \;
(t_{1n} = t_{2n}) \in \L(\G_\Q), \;
t_{1n} \neq t_{2n} }.$$
Any formula $\Q: t_1 = t_2$ in $\L(\G'_\Q)$ can then be deduced from
$\:{}1k{\forall}: t_1 = t_{1n}$ and
$\:{}1k{\forall}: t_2 = t_{2n}$ in $E'_{\forall\ldots\forall}$
and
$\Q: t_{1n} = t_{2n}$ in $E'_\Q$,
where $t_{1n}$ and $t_{2n}$ are the normal forms of $t_1$ and $t_2$,
respectively.

Finally, let $\AX_\x(\A) = \bigcup_{\Q \in \QQ(\x)} E'_\Q$.
The proof for $\AX_\x({\bf A})$ is similar.
\qed
}

Observe that the variables in $\x$ are introduced as constants into the
grammars,
hence
$E_{\forall\ldots\forall}$ in the above proof
is a set of ground equations.
A closer look at the algorithm referred by Thm.~\ref{properties}.\ref{p
externing} reveals that it generates
in fact a Noetherian ground--rewriting
system assigning unique normal forms.
Anyway, no proper instance of any formula from $\AX_\x(\A)$
is needed to derive any one in $\TH_\x(\A)$.
By permitting proper instantiations,
we may delete formulas that are instances of
others, thus reducing their total number significantly.
To find such subsumed formulas, an appropriate indexing technique may
be used, see e.g.\ \cite{Graf:1994}.

\begin{figure}
\begin{center}
\begin{tabular}{@{}l@{}}
\begin{tabular}[t]{@{}l@{\hspace*{0.5cm}}l*{10}{c}@{}}
$\forall\forall:$
& $N_{0000}$ & $:$ & \fbox{$0$}
        & $=$ & $0+0$ & $=$ & \fbox{$x+x$}
        & $=$ & $y+y$ & $=$ & $(x\.+y)+(x\.+y)$ \\
& $N_{0011}$ & $:$ & \fbox{$x$}
        & $=$ & $0+x$ & $=$ & \fbox{$x+0$}
        & $=$ & $y+(x\.+y)$ & $=$ & \fbox{$(x\.+y)+y$} \\
& $N_{0101}$ & $:$ & {$y$}
        & $=$ & $0+y$ & $=$ & $x+(x\.+y)$
        & $=$ & $y+0$ & $=$ & $(x\.+y)+x$ \\
& $N_{0110}$ & $:$ & \fbox{$x\.+y$}
        & $=$ & $0+(x\.+y)$ &     &
        & $=$ & \fbox{$y+x$} & $=$ & $(x\.+y)+0$ \\
\end{tabular}
\\
\\
\begin{tabular}[t]{@{}l*{4}{@{\hspace*{0.5cm}}ll}@{}}
$\forall\exists:$
	& $N_{0*0*}:$ & $0 = y$
	& $N_{0**0}:$ & \fbox{$0 = x\.+y$}
	& $N_{0*1*}:$ & $x = x\.+y$
	& $N_{0**1}:$ & $x = y$	\\
$\exists\forall:$
	& $N_{00**}:$ & $0 = x$
	& $N_{01**}:$ & $y = x\.+y$	\\
$\exists\exists:$
	& $N_{0***}:$ & \multicolumn{3}{l}{$0 = x = y = x\.+y$}
	& $N_{*0**}:$ & $0 = x$
	& $N_{*1**}:$ & $y = x\.+y$	\\
	& $N_{**0*}:$ & $0 = y$
	& $N_{**1*}:$ & $x = x\.+y$
	& $N_{***0}:$ & $0 = x\.+y$
	& $N_{***1}:$ & $x = y$
\end{tabular}
\\
\end{tabular}
\caption{Axioms $\AX_\x(\A)$ in Exm.~\ref{exm ax}}
\LABEL{fig exm ax}
\end{center}
\end{figure}

\EXAMPLE{}{ \LABEL{exm ax}
Continuing Exm.~\ref{exm TH},
and referring to the notions the proof of of Thm.~\ref{15},
we obtain the set $E_{\forall\forall}$ shown at the top of
Fig.~\ref{fig exm ax},
where we chose the normal form of $N_{0000}$, $N_{0011}$, $N_{0101}$, 
and
$N_{0110}$ as $0$, $x$, $y$, and $x+y$, respectively%
	\footnote{%
	The algorithm of Thm.~\ref{properties}.\ref{p externing}
	can easily be modified to work with arbitrary chosen normal 
	forms
	instead of external constants.
	},
which are each of minimal size.
From each rule alternative in Fig.~\ref{fig exm TH}, we get one equation
that is universally valid in $\A$.
The equations between marked terms remain nontrivial if instantiations
are allowed.

For each of the $N_{ijkl}$ with a ``*'', 
we check which of the above 4 normal forms
are member of $\L(N_{ijkl})$.
This can be decided quickly by matching the index;
e.g.\ $N_{0*0*}$ contains $0$ and $y$ since $0{*}0{*}$
matches with $0000$ and $0101$.
Each pair of normal forms in the same
$\L(N_{ijkl})$ gives rise to an equation, shown at the bottom of
Fig.~\ref{fig exm ax}.
Note that equations between terms of different $N_{ijkl}$ are neither in
$\TH_\x(\A)$ nor in $\AX_\x(\A)$.
For example,
``crossing'' $N_{0*0*}$ and $N_{0**0}$ yields
the forbidden formula $\forall x \exists y: y=x+y$ which
does not hold in $\A$.

After
removing all redundancies%
	\footnote{%
	E.g.\ $\exists x \forall y: y=x+y$ follows from
	$\forall x: x=x+0$ and $\forall x \forall y: x+y=y+x$.
	}%
, we get
$$\begin{array}{l*{4}{@{\hspace*{0.5cm}}rl}@{\hspace*{0.5cm}}l}
\{ 
	& \forall x: & 0=x\.+x,
	& \forall x \forall y: & x\.+y=y\.+x,
	& \forall x \exists y: & 0=x\.+y,	\\
& \forall x: & x=x\.+0,
	& \forall x \forall y: & x=(x\.+y)\.+y
	&&& \}	\\
\end{array}$$
as a set of formulas implying every closed
formula over $\set{x,y,0,(+),(=)}$ that is valid in $\A$.
Note that the associativity law is not implied, since it requires 3
variables.
\qed
}

\section{Application to Equational Theories}
\label{Application to Equational Theories}

We now show some consequences of axiomatization properties
in a purely equational setting.
Remember our convention made before Def.~\ref{19},
that in this setting, each algebra is considered to be admitted.
We restrict $\TH$ and $\AX$ to the set
$${\cal U} 
:= \set{ \forall\ldots\forall: t_1 =_S t_2 
\mid S \in \SO, \;
t_1,t_2 \in \T{\Sigma}{S}{\V} },$$
which is trivially a regular tree language.

For a given signature $\Sigma$ and a class ${\bf A}$ of
$\Sigma$--algebras, let $\vcl{\bf A}$ denote the smallest variety
containing
${\bf A}$, i.e., the class of all $\Sigma$--algebras
obtainable from algebras in ${\bf A}$ by building subalgebras, Cartesian
products, and homomorphic images.
For a set $E$ of equations, let $\Sat{E}$ denote the class of all
$\Sigma$--algebras $\A$ with $\A \sat E$.
From Birkhoff's variety theorem
\cite{Meinke:Tucker:1992},
it is well known that each
class ${\bf A}$ of algebras can
be characterized by universal equations only up to its
variety closure $\vcl{\bf A}$.
However, it is not clear in general how to find such an axiom set $E$
with $\Sat{E} = \vcl{\bf A}$.

If ${\bf A}$ is a finite class of finite algebras,
we can at least construct an increasing sequence
of tree languages
characterizing $\vcl{\bf A}$ \emph{in the limit} (Cor.~\ref{17}).
Whenever there exists any finite axiom set $E$ for $\vcl{\bf A}$ at all,
the sequence of corresponding
model classes eventually becomes stationary,
and, using Thm.~\ref{15},
we can obtain a finite axiom set that uniquely characterizes
$\vcl{\bf A}$.
However, convenient criteria for
detecting if and when the sequence becomes stationary 
are still unknown.

\COROLLARY{Variety Characterization}{ \LABEL{17}
For any finite class of finite algebras ${\bf A}$,
we can compute a sequence
$TH_1 \subseteq TH_2 \subseteq \ldots$
of sets of universal equations
such that
$\vcl{{\bf A}}
= \Sat{\bigcup_{i=1}^\infty TH_i}$.
If $\vcl{\bf A} = \Sat{E}$ for any finite $E$,
we already have $\Sat{TH_n} = \vcl{\bf A}$ for some $n \in \N$.
In this case, we can compute a finite axiom set for $\vcl{\bf A}$
from $TH_n$.
\qed
}
\PROOF{
Assuming $\V = \set{x_1,x_2,\ldots}$,
let $\x_i := \tpl{\,1n{x_\i}}$
for $i \in \N$.
Let $TH_i := \TH_{\x_i}({\bf A}) \cap {\cal U}$
and $TH_\infty := \bigcup_{i=1}^\infty TH_i$;
then, $TH_i \subseteq TH_{i+1} \subseteq TH_\infty$.
By Thm.~\ref{8},
$TH_\infty$ consists of all universal equations that hold in ${\bf A}$.

By Birkhoff's variety theorem,
$\vcl{\bf A} = \Sat{E}$ for some set $E$ of equations.
Since ${\bf A} \sat E$,
we have $E \subseteq TH_\infty$,
hence $\Sat{TH_\infty} \subseteq \Sat{E} = \vcl{\bf A}$.
Vice versa, 
we have $\vcl{\bf A} \subseteq \Sat{TH_\infty}$,
since ${\bf A} \subseteq \Sat{TH_\infty}$,
and $\Sat{TH_\infty}$ is
closed wrt.\ subalgebras, products, and homomorphic images.

If $E$ is finite, let $n \in \N$ such that all variables in $E$ occur in
$\x_n$,
then $\Sat{TH_n} = \vcl{\bf A}$ as above.
\qed
}

\section{Application to Theorem Proving}
\label{Application to Theorem Proving}

Next, we extend our results to
arbitrary formulas of first--order predicate logic.
This can easily be achieved by including a sort $Bool$ and
encoding predicates and junctors as functions to $Bool$.
We admit only algebras with an appropriate interpretation of $Bool$,
cf.\ the convention before Def.~\ref{19}.

Thus, the equation set $\TH_\x(\A)$, and $\AX_\x(\A)$ corresponds to the
set of all formulas in $\x$ valid in $\A$, and a finite
axiomatization of it, respectively.
Moreover, we can arbitrarily restrict the set of junctors that may
occur in a formula.
Note, however, that 
we cannot get rid of any equality predicate%
	\footnote{%
	Logical equivalence in $Bool$
	}%
,
as they are core components of our approach, cf.\ Def.~\ref{2}.
Hence, we cannot compute the set of all Horn formulas valid in a
given algebra.


\begin{figure}
{
\newcommand{\0}{0}
\scriptsize
\begin{tabular}[t]{@{}rcl@{}}
$x\.+x$ & $=$ & $\0$ \\
$\0\.+x$ & $=$ & $x$ \\
$x\.+\0$ & $=$ & $x$ \\
$(y\.+x)\.+y$ & $=$ & $x$ \\
$(x\.+y)\.+y$ & $=$ & $x$ \\
$(x\.+x)\.+y$ & $=$ & $y$ \\
$(x\.+y)\.+x$ & $=$ & $y$ \\
$y\.+x$ & $=$ & $x\.+y$ \\
\end{tabular}
\hfill
\begin{tabular}[t]{@{}rcl@{}}
$\0\.<x \wedge \0\.<y \wedge \0\.<x\.+y$ & $\lra$ & $false$ \\
$\0\.<x \wedge \0\.<y \wedge x\.<y$ & $\lra$ & $false$ \\
$\0\.<x \wedge \0\.<y \wedge y\.<x$ & $\lra$ & $false$ \\
$\0\.<x \wedge x\.<y$ & $\lra$ & $false$ \\
$x\.<y \wedge y\.<x$ & $\lra$ & $false$ \\
$x\.<\0$ & $\lra$ & $false$ \\
$x\.<x$ & $\lra$ & $false$ \\
$x\.+y\.<x$ & $\lra$ & $\0\.<x \wedge \0\.<y$ \\
$x\.+y\.<y$ & $\lra$ & $\0\.<x \wedge \0\.<y$ \\
$\0\.<x \wedge \0\.<x\.+y$ & $\lra$ & $y\.<x$ \\
$y\.<x \wedge (\0\.<x \vee x\.<y)$ & $\lra$ & $y\.<x$ \\
$y\.<x\.+y$ & $\lra$ & $y\.<x$ \\
\end{tabular}
\hfill
\begin{tabular}[t]{@{}rcl@{}}
$(\0\.<x \wedge \0\.<y) \vee y\.<x$ & $\lra$ & $\0\.<x$ \\
$\0\.<y \wedge \0\.<x\.+y$ & $\lra$ & $x\.<y$ \\
$x\.<x\.+y$ & $\lra$ & $x\.<y$ \\
$\0\.<y \wedge (\0\.<x \vee x\.<y)$ & $\lra$ & $\0\.<y$ \\
$(\0\.<x \wedge \0\.<y) \vee x\.<y$ & $\lra$ & $\0\.<y$ \\
$\0\.<x\.+y \wedge (\0\.<x \vee x\.<y)$ & $\lra$ & $\0\.<x\.+y$ \\
$x\.<y \vee y\.<x$ & $\lra$ & $\0\.<x\.+y$ \\
$(\0\.<x \wedge \0\.<y) \vee \0\.<x\.+y$ & $\lra$
	& $\0\.<x \vee x\.<y$ \\
$\0\.<x \vee \0\.<y$ & $\lra$ & $\0\.<x \vee x\.<y$ \\
$\0\.<x \vee \0\.<x\.+y$ & $\lra$ & $\0\.<x \vee x\.<y$ \\
$\0\.<y \vee \0\.<x\.+y$ & $\lra$ & $\0\.<x \vee x\.<y$ \\
$\0\.<y \vee y\.<x$ & $\lra$ & $\0\.<x \vee x\.<y$ \\
\end{tabular}

}
\caption{Axiom Set in Exm.\ \protect\ref{18}}
\label{Axiom Set in Exm. 18}
\end{figure}

\EXAMPLE{}{ \LABEL{18}
As an example of computed predicate--logic axiomatizations,
consider $(\N \;mod\; 2)$ with one function $(+)$
and one predicate $(<)$, where
the sort $Bool$ is interpreted by the two--element Boolean algebra, as
required.

Any universally quantified formula in the variables $x,y$ and
with $\wedge$ and $\vee$ as the only logical junctors
that holds in $(\N \;mod\; 2)$
follows from the set of formulas given in Fig.~\ref{Axiom Set in Exm.
18};
formulas with other quantifier prefixes are omitted.

Pure ground formulas and formulas that are instances
of others have been manually
deleted, as well as formulas that follow from
propositional tautologies or symmetry of equality.
Equations in $Nat$ are listed first, with index $_{Nat}$ omitted,
followed by equations in $Bool$, $=_{Bool}$ written as $\lra$.
Note that $(=_{Nat})$ 
is {\em not} contained in the signature of this example
but was introduced by our method;
consequently, no equality predicate appears in the equations of
sort $Bool$,
e.g.\ in a law like $x<y \vee x=y \vee y<x \lra true$.

No formula at all can be reduced to $true$ by the axioms in
Fig.~\ref{Axiom Set in Exm. 18}, indicating that no
valid statements about $(\N\;mod\;2)$ can be expressed
in $\Sigma$,
except for trivial propositional instances like $true \vee x<y$.
In fact, every expressible formula (i.e.\ term
in $\T{\Sigma}{Bool}{\set{x,y}}$)
can be falsified by instantiating
both $x$ and $y$ to $0$.
\qed
}

In Cor.~\ref{20},
we give an application in the field of
automated theorem proving.
Here, it is common practice to test a conjecture $\phi$
in a finite class ${\bf A}$ of models
of the background theory $\Theta$
before attempting to prove $\phi$ from $\Theta$.
If the test fails, it is clear that $\Theta \impl \phi$ cannot hold.
If the test succeeds, i.e.\ ${\bf A} \sat \phi$,
we are usually still faced with
the task of proving $\Theta \impl \phi$.

We call ${\bf A}$ a class of prototype
algebras for $\Theta$, if from a succeeding test we always can conclude
the validity of $\Theta \impl \phi$.
In this case, we can decide quickly
whether $\Theta$ entails a formula
$\phi$, merely by testing whether $\phi$ holds in each member of
${\bf A}$.

Corollary~\ref{20} provides a sufficient criterion for establishing
the
existence of prototype algebras for an equational
--~or by the above argument~--
first--order predicate--logic theory $\Theta$.
Since we cannot deal with arbitrarily
many variables, we have to restrict
the syntactic class of $\phi$ to 
$\T{\Sigma \cup \set{(=_{Bool})}}{}{\x}$ for some
finite tuple $\x$ of variables.

Example~\ref{21} gives a set of
prototype algebras for an equational theory,
and at the same time
shows that they do not exist for arbitrary
theories.
It remains to be seen whether it is feasible to extend the
prototype approach by adding certain infinite algebras that allow
easy testing%
	\footnote{%
	In an equational setting, the singleton class containing
	the initial algebra is always a prototype.
	However, equality in the initial algebra is generally
	undecidable.
	}
of $\phi$ to ${\bf A}$.

\COROLLARY{Prototype algebras}{ \LABEL{20}
Let $\Theta$ be a set of formulas,
let ${\bf A}$ be a finite class of finite admitted $\Sigma$--algebras
such that ${\bf A} \sat \Theta$ 
and $\Theta \impl \AX_\x({\bf A})$.
Then, ${\bf A}$ is a class of
prototype algebras for $\Theta$ and $\x$.
Formally,
for any formula $\phi$ over $\Sigma$ and $\x$
we have
$\Theta \impl \phi$
iff
${\bf A} \sat \phi$.
}
\PROOF{
$\;$\\
``$\Ra$'':
~
trivial: ${\bf A} \sat \Theta \impl \phi$
\\
``$\La$'':
~
\begin{tabular}[t]{@{}lrll@{\hspace*{1.0cm}}l@{}}
& ${\bf A}$ & $\sat$ & $\phi$	\\
$\Ra$ & $\AX_V({\bf A})$ & $\impl$ & $\phi$
	& by Thm.~\ref{15}	\\
$\Ra$ & $\Theta$ & $\impl$ & $\phi$
	& since $\Theta \impl \AX_V({\bf A})$	\\
\end{tabular}
\\
Note that the Corollary holds for arbitrary $\SO_{\fix}$ and
$\A_{\fix}$, not only for $Bool$.
However, we need to fix $Bool$ in order to get the notion of $\impl$ 
used
in theorem proving.
\qed
}

\EXAMPLE{}{ \LABEL{21}
Let $\Theta$ consist of the axioms for an Abelian group of 
characteristic~2.
Referring to Exm.~\ref{exm ax},
we can see that $\Theta$ implies $\AX_{\tpl{x,y}}(\A_2)$.
Hence, in order to prove a formula $\phi$ over $\set{x,y,0,(+),(=)}$
to be a consequence of $\Theta$, it is sufficient
by Cor.~\ref{20} just to test it in $\A_2$.

Unfortunately, we cannot get rid of ``characteristic'' equations:
In any finite algebra with an associative binary operation $(+)$,
a law $n_1x = n_2x$
holds for some $n_1 > n_2$,
where we abbreviate $nx := x+\ldots+x$ ($n$ times).
Hence, each axiom set obtained from finitely
many of such algebras necessarily entails a law of this form.
\qed
}

\small

\paragraph{Acknowledgments.}
Ingo Dahn drew our attention to the application area of prototype
algebras in automated theorem proving.
Martin Simons provided the literature
reference to the many--sorted version
of Birkhoff's variety theorem.
Angela Sodan gave us some valuable advice on the presentation.

\bibliographystyle{alpha}
\bibliography{lit}

\end{document}